\documentclass[fleqn,usenatbib]{mnras}

\usepackage{newtxtext,newtxmath}

\usepackage[T1]{fontenc}

\DeclareRobustCommand{\VAN}[3]{#2}
\let\VANthebibliography\thebibliography
\def\thebibliography{\DeclareRobustCommand{\VAN}[3]{##3}\VANthebibliography}

\usepackage{caption}
\usepackage{subcaption}
\usepackage{graphicx}	
\usepackage{amsmath}	
\usepackage{xcolor}

\newcommand{\kms}{km~s$^{-1}$}

\title[Galactic Underworld]{The Galactic Underworld: The spatial distribution of compact remnants}

\author[D. Sweeney et al.]{
David Sweeney,$^{1}$\thanks{E-mail: david.sweeney@sydney.edu.au}
Peter Tuthill,$^{1}$
Sanjib Sharma,$^{1}$
Ryosuke Hirai$^{2,3}$
\\
$^{1}$Sydney Institute for Astronomy (SIfA), The University of Sydney, Physics Road, Sydney 2006, Australia\\
$^{2}$OzGrav: The Australian Research Council Centre of Excellence for Gravitational Wave Discovery, Clayton, VIC 3800, Australia\\
$^{3}$School of Physics and Astronomy, Monash University, VIC 3800, Australia
}

\date{Accepted XXX. Received YYY; in original form ZZZ}

\pubyear{2022}

\begin{document}
\label{firstpage}
\pagerange{\pageref{firstpage}--\pageref{lastpage}}
\maketitle

\begin{abstract}
We chart the expected Galactic distribution of neutron stars and black holes. These compact remnants of dead stars --- the \emph{Galactic underworld} --- are found to exhibit a fundamentally different distribution and structure to the visible Galaxy.
Compared to the visible Galaxy, concentration into a thin flattened disk structure is much less evident with the scale height more than tripling to $1260 \pm 30$~pc. 
This difference arises from two primary causes.
Firstly, the distribution is in part inherited from the integration over the evolving structure of the Galaxy itself (and hence the changing distribution of the parent stars).
Secondly, an even larger effect arises from the natal kick received by the remnant at the event of its supernova birth.
Due to this kick we find 30\% of remnants have sufficient kinetic energy to entirely escape the Galactic potential (40\% of neutron stars and 2\% of black holes) leading to a Galactic mass loss integrated to the present day of $\sim 0.4\%$ of the stellar mass of the Galaxy.
The black hole -- neutron star fraction increases near the Galactic centre: a consequence of smaller kick velocities in the former.
(the assumption made is that kick velocity is inversely proportional to mass).
Our simulated remnant distribution yields probable distances of 19~pc and 21~pc to the nearest neutron star and black hole respectively, while our nearest probable magnetar lies at 4.2~kpc.
Although the underworld only contains of order $\sim 1\% $ of the Galaxy's mass, observational signatures and physical traces of its population, such as microlensing, will become increasingly present in data ranging from gravitational wave detectors to high precision surveys from space missions such as Gaia.
\end{abstract}




\section{Introduction}
The expected Galactic distribution of compact supernova remnants --- neutron stars (NSs) and black holes (BHs) --- has not been well established. NSs and BHs are formed when massive stars \citep[$\gtrapprox$ 8 solar masses, M$_{\sun}$;][]{smartt2009, burrows2012, postnov2014, smartt2015, muller2016}  either undergo core-collapse supernovae or direct collapse \citep{Fryer2012} at the end of their life cycle.
Traditionally, stars larger than 25~M$_{\sun}$ were thought to collapse into BHs \citep{Heger2003}.

NSs are mostly discovered by radio telescopes \citep{2004LorimerBook}, and some of them have high accuracy astrometry measured through Very Long Baseline Interferometry (VLBI) \citep{Verbunt2017} producing a sample restricted to radio-luminous pulsars. There are also many pulsars discovered in binaries, emitting X-rays powered by accretion from the companion star \citep{reig2011}. Some young isolated NSs are also discovered in X-rays, such as X-ray dim isolated NSs, central compact objects of supernova remnants, anomalous X-ray pulsars and soft gamma repeaters \citep[]{Mereghetti2011}. 
On the other hand, prior to the advent of gravitational-wave interferometers, stellar mass BHs were only observed when they were a component of an X-ray binary and are consequently rare \citep{ozel2010, spera2015}. 
Since the initial detection of gravitational waves with the LIGO and Virgo instruments, witnessing binary BH-BH, BH-NS and NS-NS mergers has opened a new window on the cosmic population of binary remnants \citep{LIGO2017, LIGO2019, LIGO2020, LIGO2021}.

The supernova event marking the birth of both NSs and BHs also injects a dynamical perturbation onto the population of remnants.
This can arise in two ways. 
Firstly, they receive a significant natal kick \citep{Lyne1994, Barack2019} from the asymmetry in the explosion and secondly, if occurring in a binary \citep[which are prevalent among massive stars;][]{Sana2012,Duchene2013}, the sudden mass loss can sometimes disrupt the orbit, ejecting the components with velocities arising from their former orbital angular momentum \citep{Blaauw1961}.
As a net result of these effects we expect a significant alteration of the Galactic distribution of remnants compared to that of their progenitor stars \citep{Lyne1994, Repetto2012}. 

The magnitude of natal kicks has long been an area of significant study \citep{Arnett1987} with some researchers attempting to derive this from hydrodynamical simulations \citep{Herant1992, Janka1994, Janka2012, Mandel2020} --- see the review by \cite{Muller2020} for a detailed overview --- while others infer it from observations \citep{Lyne1994, Hansen1997, Arzoumanian2002, Hobbs2005, Giguere2006, Verbunt2017, Katsuda2018, Igoshev2020, Igoshev2021}. 
The order of magnitude of such kicks is normally observed to lie in the 100's~\kms, with more extreme events ranging up to >1\,000~\kms.


A number of previous studies have produced simulations of subsections of the Galactic remnant distribution. \cite{Laberts2018} focus only on BH-BH binaries, \cite{Vigna2018} focus only on NS-NS binaries and \cite{Olejak2020} only model BHs. Of these, the best comparison to our work is the study by \cite{Olejak2020}. The main highlight of their work is that they follow the stellar evolution of both single and binary systems by using the code StarTrack \citep{StarTrack}. 
However, none of these previous works have explored the {\it spatial distribution of remnants} in the Galaxy. For example, the vertical distribution of disc stars in \cite{Olejak2020} is a simple geometrical approximation: a uniform density with a thickness of 0.3 kpc. 
Furthermore, \cite{Olejak2020} do not evolve natal kicks through the Galactic potential, which plays an important role in shaping the current-day distribution. 
Here we present a next-generation model with sophistication to overcome both these shortcomings. We use a population synthesis model that is state of the art in terms of modelling the spatial and age distribution of stars and evolve the remnants over cosmic time in the potential of the Galaxy to the present day.

With the first detection of a stellar mass BH via microlensing \citep{Lam2022, Sahu2022}, it is becoming increasingly important to chart the distribution of the Galactic underworld to maximise the efficiency of future searches. Furthermore, there are many more ambiguous detections and previous studies have suggested that hundreds of compact objects may be discovered through microlensing \citep[e.g.][]{Wyrzykowski2016, Wyrzykowski2020}. Furthermore, other techniques are being developed to view isolated BHs \citep{Matsumoto2018, Kimura2021} with the paper by \cite{Kimura2021} focusing on prospects for identifying nearby ($\lesssim1$~kpc) isolated BHs. Hence, a detailed understanding of the spatial and kinematic distribution of the Galactic underworld --- BHs and NSs --- is both timely and relevant.
Providing such a map of the galactic underworld is the motivation for this work.


\section{Methods}
\label{sec:methods}
To model the Milky Way stellar distribution, we use the stellar population synthesis code {\tt GALAXIA} \citep{Galaxia}.
{\tt GALAXIA} generates a synthetic catalog of stars with position (${\bf x}$), velocity (${\bf v}$), age (${\tau}$), metallicity ([M/H]), mass ($m$), stellar parameters and photometry by sampling stars from a prescribed theoretical model of the Galaxy. 
The Galactic model used is based on the well established Besancon Galaxy model of \citet{2003A&A...409..523R} 
but with some modifications. The Besancon Galaxy model has been tested against stars counts from photometric surveys and is built to satisfy a number of observational constraints such as velocity dispersion as a function of age.
The Galactic model specifies the number density distribution of stars in the space of $(x,v,\tau,\mathrm{[M/H]}, m)$. Theoretical stellar isochrones are used to compute stellar parameters and photometry from $\tau, m$ and [M/H]. 
{\tt GALAXIA} models the 
Milky Way as a superposition of 
four distinct Galactic components: a thin disc, a thick disc, a stellar halo and a triaxial bulge.
Analytical functions are used to model the density distributions ($p(x|\tau)$), initial mass function (IMF) ($p(m)$), star formation rate ($p(\tau)$) and metallicity as a function of age $\big(p(\mathrm{[M/H]}|\tau)\big)$ for the different components. The IMF and star formation/density normalisations for the 
different Galactic components are listed in \autoref{tab:gmodels1}. 
The values for the thin disc and bulge 
were taken from \cite{Galaxia}. 
However, for the thick disc and the stellar halo, $\alpha_2$ was changed from $-0.5$ to $-2.35$, as the former was too shallow as compared to most other studies (see \cite{Hopkins2018} and references therein). 
The local density normalisations were adjusted so that the number of visible stars today for these components (which were typically less than solar mass) remained approximately the same as in \cite{Galaxia}.

The isochrones used in {\tt GALAXIA} to predict the stellar properties  are from the Padova database using CMD 3.0 (\href{http://stev.oapd.inaf.it/cmd}{http://stev.oapd.inaf.it/cmd}), with  PARSEC-v1.2S isochrones \citep{Bressan2012, Tang2014, Chen2014, Chen2015}, the NBC version of bolometric corrections \citep{Chen2014}, and assuming Reimers mass loss with efficiency $\eta=0.2$ for RGB stars. The isochrones are computed for scaled-solar composition following the $Y=0.2485+1.78Z$ relation and their solar metal content is $Z_{\odot}=0.0152$. The same isochrones were also used to estimate the stellar lifetimes.

By default {\tt GALAXIA} produces live stars burning nuclear fuel. For the purpose of this paper, we modified {\tt GALAXIA} to also output stars that exhaust their nuclear fusion life cycle, leaving behind a remnant.
The initial stellar mass was used to decide the star's fate and the kind of remnant generated: a NS (8 to $25 {\rm M_{\odot}}$) or a BH ($> 25 {\rm M_{\odot}}$). Stars with initial mass <~8~M$_\odot$ were deemed insufficiently massive to form into either NSs or BHs and were filtered out of the {\tt GALAXIA} output dataset.

Recent studies suggest that the boundary between NS and BHs may not be so clear cut.
A detailed understanding of the many physical pathways leading to the creation of compact remnants from given progenitor stellar properties is an active area of research \citep{Ugliano2012, Ertl2016, Sukhbold2016, Mandel2020}. 
To circumvent this complexity, many works elect to split stellar remnants by progenitor mass \citep{Fryer1999, Belczynski2008, Fryer2012, Roman2021, Schneider2021}.
There is some observational and theoretical support for this division: \citet{smartt2015} observed 18 supernova progenitors $\lesssim$~18~M$_{\sun}$ and found that all collapsed into NSs. 
Furthermore, \citet{Adams2017} identified a red supergiant of mass $\sim$25~M$_{\sun}$ which vanished, implying a direct collapse into a BH.
With this in mind, 8--25~M$_{\sun}$ is a typical range for initial masses of NS progenitors and >25~M$_{\sun}$ for initial masses of BH progenitors.

\begin{figure}
	\includegraphics[width=\columnwidth]{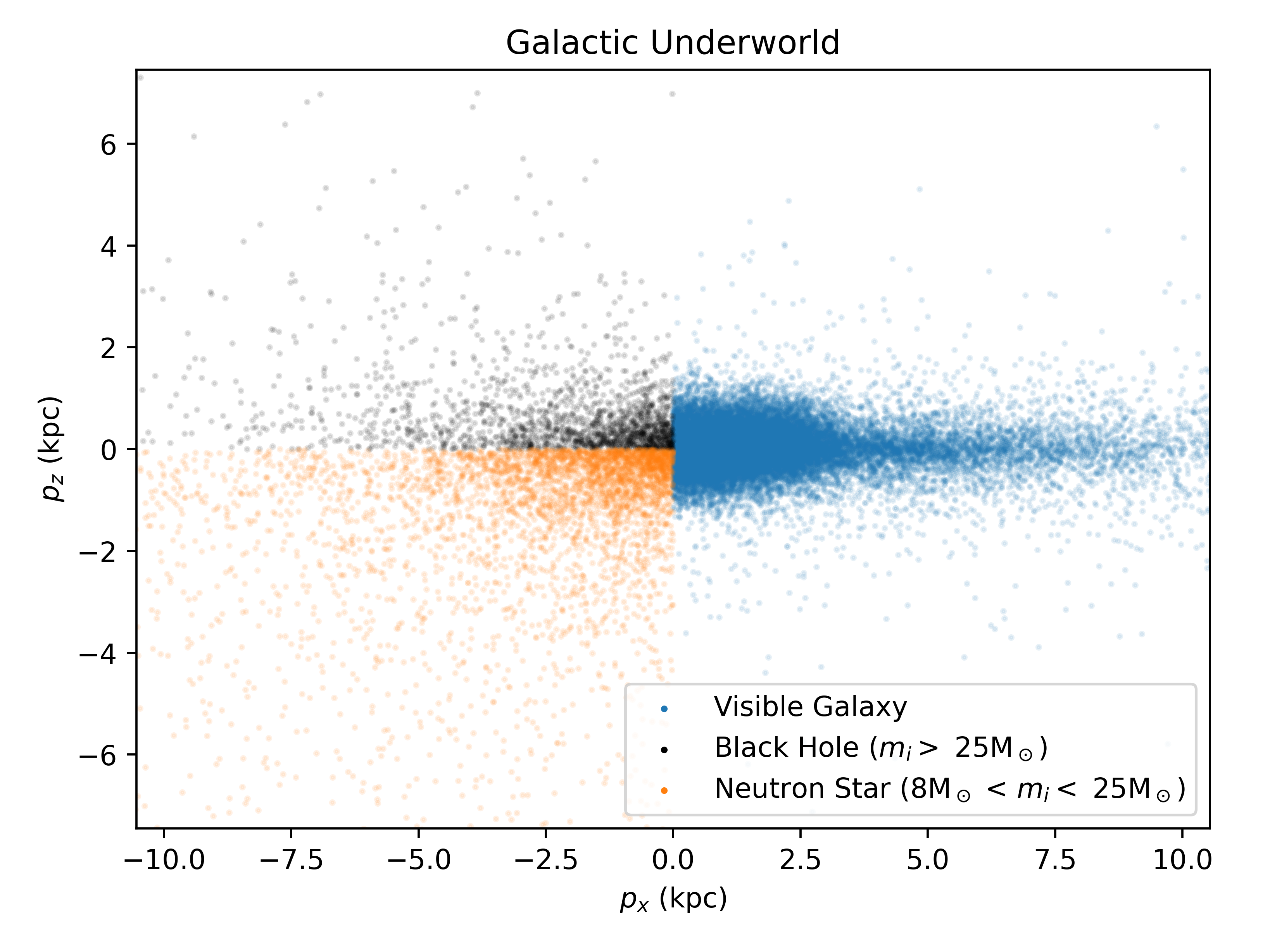}
    \caption{Depiction of the spatial distribution of objects generated by {\tt GALAXIA} once natal kicks were accounted for using {\tt galpy}. The visible Galaxy and the Galactic underworld are shown in a side-by-side comparison with BHs only plotted onto the top-left quadrant and NSs only plotted onto the bottom-left quadrant. $m_i$ is the initial mass of the star.}
    \label{fig:kicked_data}
\end{figure}

\begin{table}
\caption{The IMFs and the density normalisations of Galactic components in terms of star forming mass. IMF is normalised to have mean mass of 1 ${\rm M_{\odot}}$ in range  of 0.07 to 100 $M_{\odot}$.
The parameters $\alpha_1$ and $\alpha_2$ are used to specify the IMF (number density of stars as a function of stellar mass, $M$), which is of the following form, $\propto M^{\alpha_1}$ for $M/{\rm M}_{\odot}<1$ and $\propto M^{\alpha_2}$ for $M/{\rm M}_{\odot} >1$. }
\begin{tabular}{l l l l}
\hline
Galactic Component & Normalisation & $\alpha_1$  & $\alpha_2$\\
\hline
Thin ($0\text{--}7$~Gyr) & \textsuperscript{a}
2.37 
${\rm M_{\odot} yr^{-1}}$ 
& -1.6 & -3.0 \\
Thin ($7\text{--}10$~Gyr)& \textsuperscript{a} 
$1.896 {\rm M_{\odot} yr^{-1}}$ 
& -1.6 & -3.0 \\
Thick (11 Gyr)& 
\textsuperscript{b} 
$6.0077 \times 10^6 {\rm M_{\odot} kpc^{-3}}$ & -0.5 & -2.35 \\
Stellar Halo (13 Gyr) & \textsuperscript{b}
$6.5776 \times 10^{4} {\rm M_{\odot}kpc^{-3}}$ & -0.5 & -2.35 \\
Bulge (10 Gyr) & 
\textsuperscript{c}
$3.5088 \times 10^{9} {\rm M_{\odot}kpc^{-3}}$
& -2.35 & -2.35 \\
\hline
\multicolumn{4}{l}{
\textsuperscript{a}{Star formation rate}} \\
\multicolumn{4}{l}{\textsuperscript{b}{Local mass density}} \\
\multicolumn{4}{l}{\textsuperscript{c}{Central density}}\\
\end{tabular}
\label{tab:gmodels1}
\end{table}

Stars which experience supernovae, and therefore a natal kick, have their velocity significantly altered. This mechanism is not captured in {\tt GALAXIA}'s remnant distribution and so was modelled in a custom code written for this purpose. The speed distribution imparted by kicks is found to be bimodal with a low-velocity peak, attributed to electron-capture supernovae, and a high-velocity peak, attributed to standard iron core-collapse supernovae \citep{Beniamini2016, Verbunt2017, Vigna2018, Igoshev2020, Igoshev2021}. \cite{Igoshev2020} find the distribution of pulsars to be a bimodal Maxwellian given by:
\begin{align}
    f(v | w, \sigma_1, \sigma_2) \mathrm{d}v = w \mathcal{M}(v | \sigma_1) \mathrm{d}v + (1 - w) \mathcal{M}(v | \sigma_2) \mathrm{d}v \label{eq:bimodal}
\end{align}
Where $v$ is the magnitude of the natal kick, $\mathcal{M}$ is a maxwellian distribution and $w, \sigma_1$ and $\sigma_2$ are parameters which have values of $w = 0.2, \sigma_1 = 56$~km/s and $\sigma_2 = 336$~km/s. These parameters are not significantly different to the later work by \cite{Igoshev2021} which was published while our modelling was progressing. The pulsars selected in \citet{Igoshev2020} are all single pulsars. Since it is not trivial to trace back whether a pulsar was previously in a binary or not, the sample likely contains a significant fraction of NSs that escaped from a binary companion. Therefore, our chosen kick distribution implicitly accounts for effects related to binarity as long as we here focus on pulsars that are currently single.

Natal kicks were assumed to impart the same momentum regardless of whether the remnant is a BH or NS. 
Expected remnant masses will of course exhibit significant variance carried from the diversity of progenitor and evolutionary pathway. 
However, widely accepted values of about 1.35~M$_{\sun}$ for NSs \citep{ozel2012, postnov2014, Sukhbold2016} and 7.8~M$_{\sun}$ for BHs \citep{ozel2010, spera2015, Sukhbold2016}, when taken in ratio,  offer an approximation to model the BH population.
This leads to natal kicks having a magnitude sampled from the following formula and being oriented in a random 3D orientation: 
\begin{align} \label{eq:kick}
    f(v | w, \sigma_1, \sigma_2,& m_{\text NS}, m_{\text R}) \mathrm{d}v \nonumber\\
    &= \frac{m_{\text NS}}{m_{\text R}}\left[ w \mathcal{M}(v | \sigma_1) + (1 - w) \mathcal{M}(v | \sigma_2) \right ] \mathrm{d}v
\end{align}
Where $m_{\text NS}$ is the mass of a typical NS (1.35~M$_\odot$) and $m_{\text R}$ is the mass of a typical NS (if sampling for NSs) or BH (7.8~M$_\odot$; if sampling for BHs). This results in BHs gaining a smaller natal kick ($5.8$ times smaller) as their remnant mass is significantly larger than a NS remnant. 
BH kicks are an active area of study, so other works often make this simplifying assumption \citep{Whalen2012, Antonini2016}.
We note that there is some circumstantial evidence suggesting that BHs receive larger kicks than those we provide \citep{Repetto2012, Repetto2017, Vanbeveren2020}. BHs with progenitors massive enough for direct collapse (> 40~M$_\odot$) had their kick set to 0~km/s.

These kicks, assumed to be directed isotropically with respect to the progenitor, were then added to each remnant's velocity provided by {\tt GALAXIA} (transformed to galactocentric coordinates). The remnants were then evolved using {\tt galpy} \citep{galpy} since the formation of the remnant until today (age of the progenitor minus the lifetime of the progenitor) in the {\tt MWPotential2014} Galactic potential with solar radius parameter $R_{\sun} = 8$~kpc \citep{1993ARA&A..31..345R} and circular velocity at Sun $V_{\rm circ}(R_{\sun}) = 232$~km/s \citep{2014ApJ...793...51S}.

\section{Results and Discussion}
\label{sec:results}
\subsection{Spatial distribution of remnants}
\begin{figure*}
    \centering
    \begin{subfigure}[t]{\columnwidth}
        \centering   	\includegraphics[width=\columnwidth]{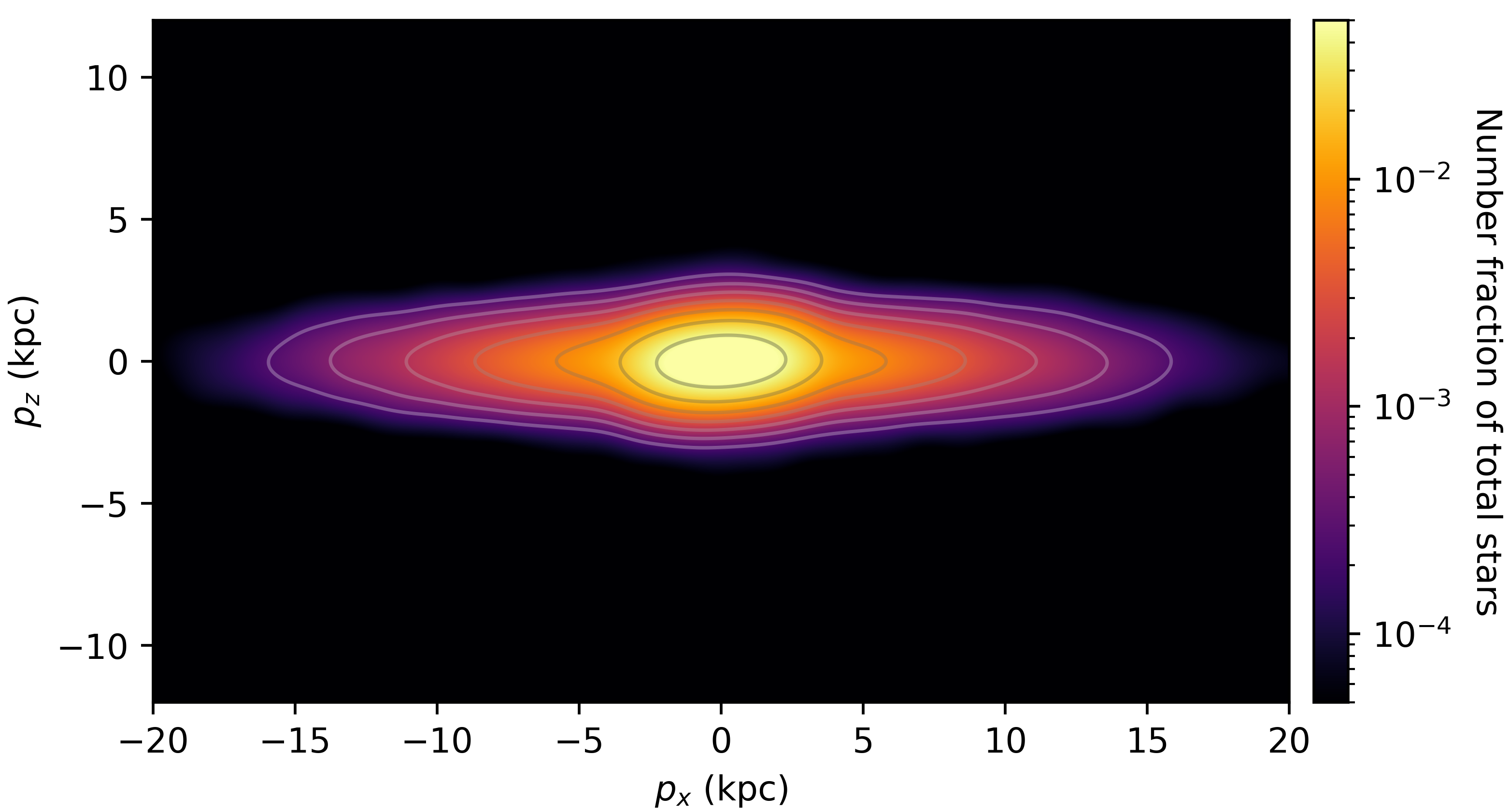}
    	\caption{Visible Galaxy rendered from {\tt GALAXIA}}\label{fig:visible}
    \end{subfigure}
    \hfill
    \begin{subfigure}[t]{\columnwidth}
        \centering
    	\includegraphics[width=\columnwidth]{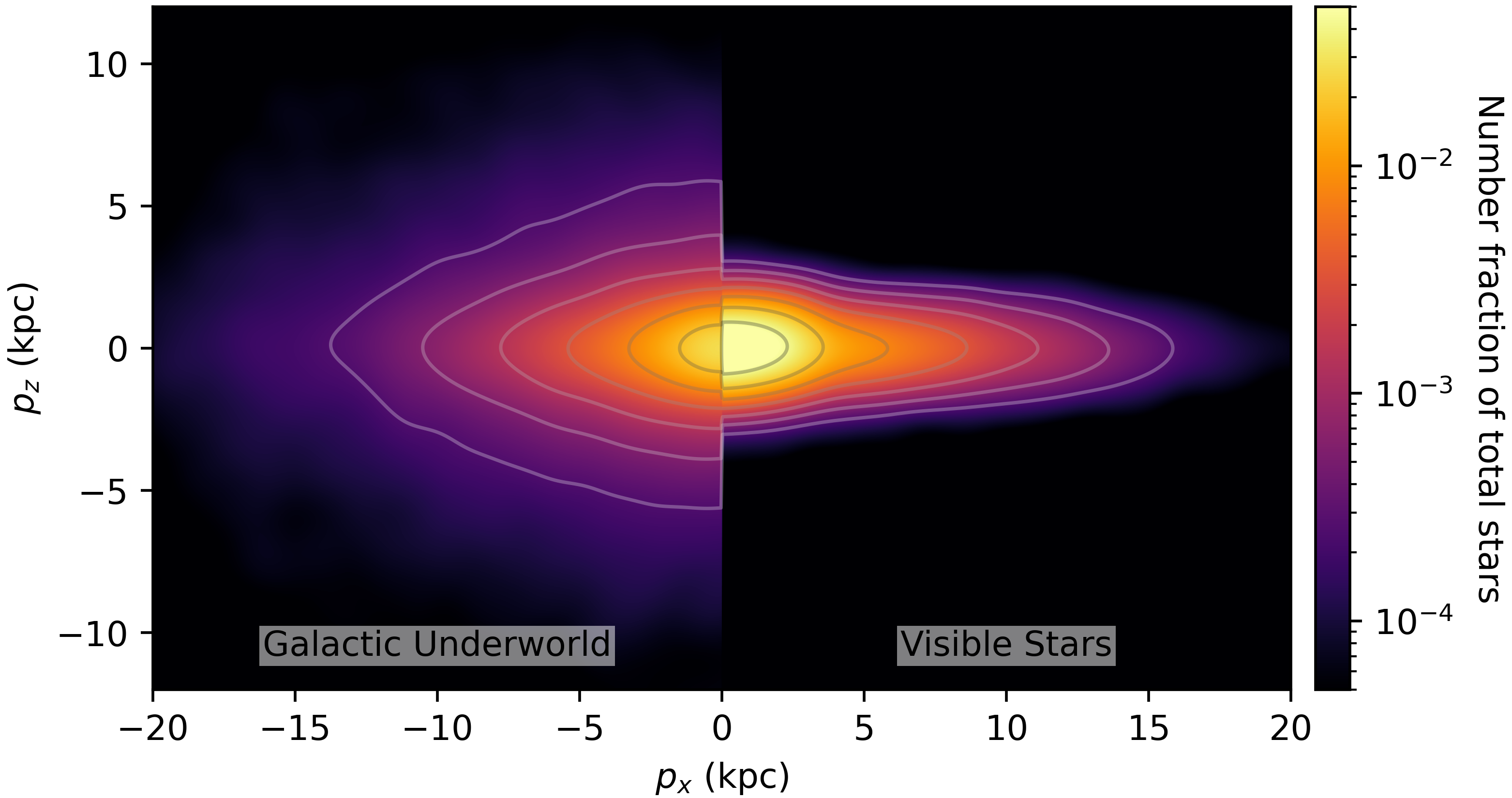}
    	\caption{Galactic underworld (left side) with visible Galaxy (right side)}\label{fig:guw}
    \end{subfigure}
    
    \begin{subfigure}[t]{\columnwidth}
        \centering
    	\includegraphics[width=\columnwidth]{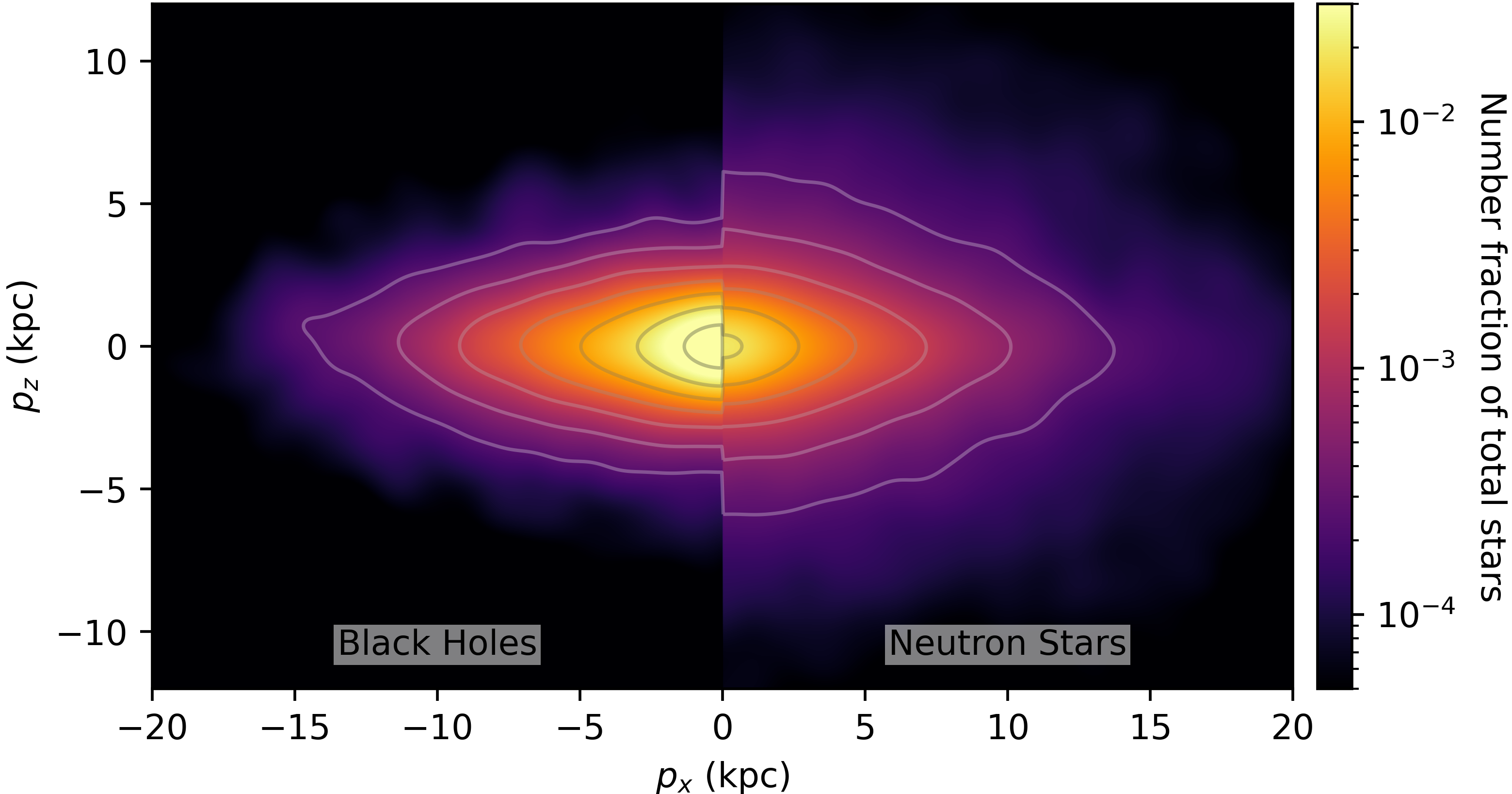}
    	\caption{Black holes (left side) with neutron stars (right side)}\label{fig:type}
    \end{subfigure}
    \hfill
    \begin{subfigure}[t]{\columnwidth}
        \centering
    	\includegraphics[width=\columnwidth]{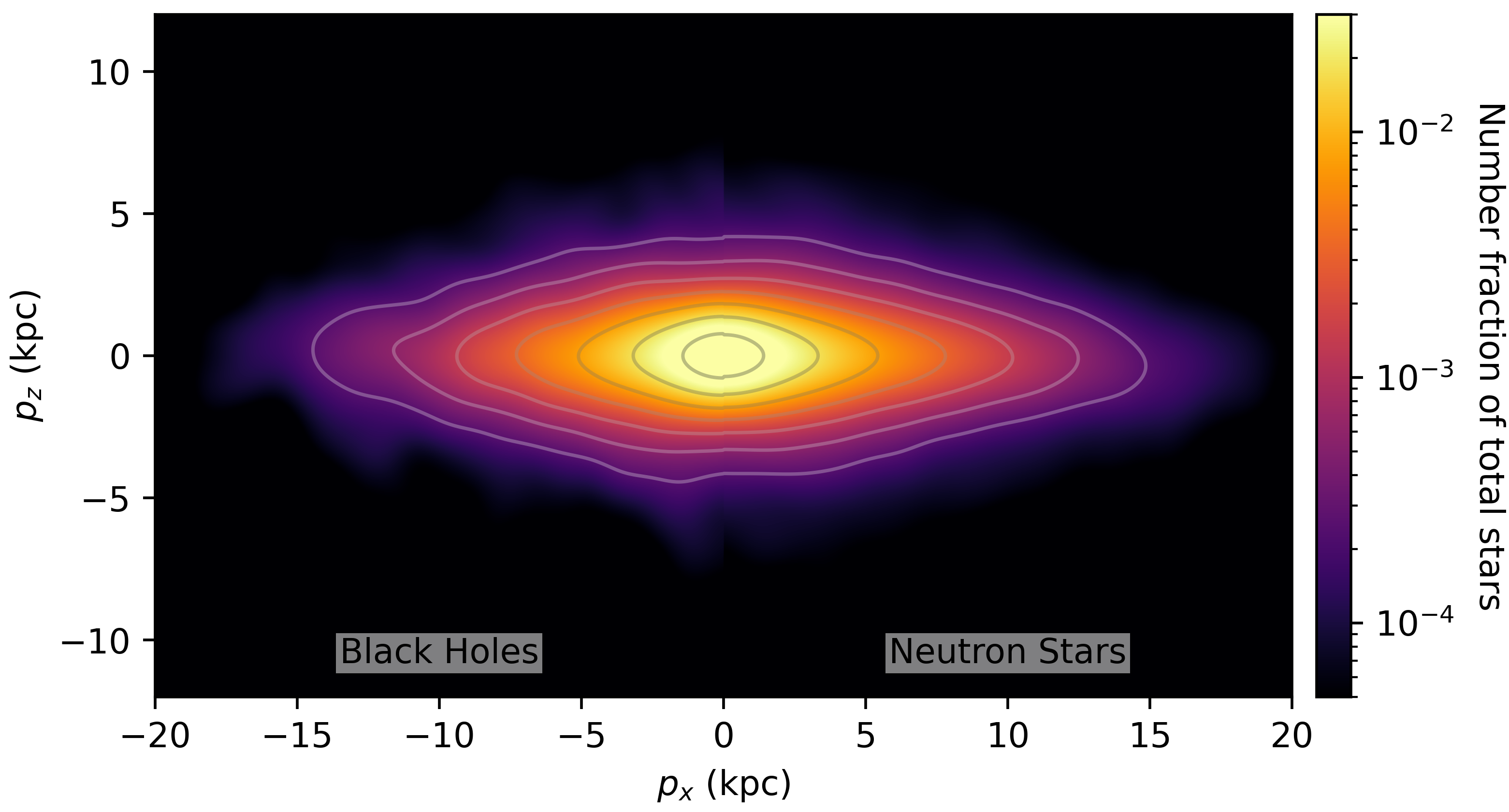}
    	\caption{Undisrupted massive binaries: BH binaries (left side) with NS binaries (right side)}\label{fig:renzo}
    \end{subfigure}
    \caption{Probability density map of visible stars and the Galactic underworld in the Galactocentric $x-z$ plane.  Logarithmically spaced contour lines are plotted on top of all plots.
    Panel (\subref{fig:visible}): visible Galaxy generated by {\tt GALAXIA}. Panel (\subref{fig:guw}): Galactic underworld plotted side-by-side the visible Galaxy. Panel (\subref{fig:type}): side-by-side distributions of NSs and BHs, it is clear that BHs are much more likely to be located near the Galactic centre, with NSs much more dispersed. Panel (\subref{fig:renzo}): side-by-side distributions of NS and BH binaries. These plots were generated by passing the star/remnant locations through a gaussian kernel density estimator \citep{rosenblatt1956, parzen1962}. The colours indicate probability density (fraction of stars per unit area) such that its  integral over the whole $x-z$ plane is 1.}
    \label{fig:contours}
\end{figure*}


A scatter plot of the Galactic underworld resulting from our simulation, together with the normal visible star Galaxy for comparison, is depicted in \autoref{fig:kicked_data}.
Even to a casual inspection it is immediately apparent that the Galactic underworld has a fundamentally different shape and structure relative to the visible Galaxy. 
We illustrate this more clearly in \autoref{fig:guw}, which shows the probability density map in the Galactic $x$--$z$ plane of stellar remnants alongside that of stars in the visible Galaxy.
We can see that the vertical axis of the distribution has puffed up with the result more closely resembling a spherical cloud than the thin disk of the Milky Way. 
This can be understood by considering that even a small kick for a remnant $10$~kpc away from the Galactic centre will result in the remnant either acquiring escape velocity, or being boosted to a sufficiently eccentric orbit as to no longer keep station in the Galactic disk.

These kicks perturbing the remnants out of the Galactic disk cause the Galactic underworld to be much more diffuse than the visible Galaxy. 
As shown in \autoref{tab:scale_heights}, {\tt GALAXIA}'s synthetic visible Galaxy has a scale height of $334 \pm 8$~pc and a scale length, calculated up to the 50th percentile of radius, of $920 \pm 30$~pc while the Galactic underworld has a scale height of $1260 \pm 30$~pc and a scale length, up the the 50th percentile, of $860 \pm 20$~pc. 
Calculating the scale length up to the 50th percentile of radius was chosen because of increasing uncertainty for larger radii and stars being best fit by a broken power law, as shown in \autoref{fig:log_marginal_dist}.
The thicker disk of the early Galaxy (in which many progenitors arose) together with natal kicks acts to transform the Galactic underworld to be less concentrated in the plane and bulge; a feature which is depicted in \autoref{fig:guw}. The scale height of a relaxed pulsar population is expected to be at least 500~pc \citep{Lorimer2008}. 
Defining relaxed pulsars as NSs with an age between 10~Myr and 100~Myr, we find the scale height of this population to be $630 \pm 70$~pc, in agreement with the value from \citet{Lorimer2008}.
At the location of the Sun, this difference in structure results in our local space density of remnants being 1 remnant per $1.7\times 10^4$~pc$^3$, or a probable distance to the nearest remnant of 16~pc (19~pc for a NS and 21~pc for a BH).
We calculate the probable distance to the nearest young NS (<~100~Myr old) to be 50~pc; for comparison the nearest known pulsars are 100--200 pc away \citep{ANTF}. These space densities were calculated by totalling the remnants found within a torus, in the plane of the Galaxy, with a major radius of 8~kpc (corresponding to the Sun's location) and a minor radius of 200~pc.

\begin{table}
	\centering
	\caption{Comparison of the scale heights and lengths of the the visible Galaxy, the Galactic underworld without the effects of natal kicks, the Galactic underworld (all remnants with natal kicks) and then of NSs and BHs individually. The scale lengths are calculated on the inner 50\% of stars/remnants (Galactocentric cylindrical radii $\lesssim 2.5$~kpc) due to increasing uncertainty at larger radii and stars being best fit with a broken power law, see \autoref{fig:log_marginal_dist}. The provided uncertainty is one standard deviation.}
	\label{tab:scale_heights}
	\begin{tabular}{lcc}
		\hline
		 & Scale Height (pc) & \shortstack{Scale Length (pc)\\(50 percentile)}\\
		\hline
		Visible Galaxy & $334 \pm 8$ &  $920 \pm 30$\\
		Unkicked Galactic Underworld & $560 \pm 10$ & $930 \pm 20$ \\
		Galactic Underworld & $1260 \pm 30$ & $860 \pm 20$ \\
		Neutron Stars & $1490 \pm 50$ & $950 \pm 20$ \\
		Black Holes & $900 \pm 40$ & $750 \pm 20$ \\
		\hline
	\end{tabular}
\end{table}

The larger masses of BH remnants attenuates the velocities imparted by the natal kicks, leading to more BHs being retained in the disk; moreover these are found closer to the Galactic centre as compared to the NS population. 
This difference in remnant distribution can be seen in \autoref{fig:type}. 
The scale heights and lengths of NSs and BHs, compared to those of the visible Galaxy, are shown in \autoref{tab:scale_heights}. 
Plots of marginal distributions as a function of radius and height can be found in Appendix~\ref{sec:marginal_dist}. 

As the LIGO/Virgo gravitational wave interferometers accumulate events, with sufficiently rich observational support it may eventually become possible to place constraints on the distribution of mergers with location in the host Galaxy population. 
However, such inspiral mergers are drawn from a different population to the one studied in this work.
While the velocity distribution of pulsars is mostly unaffected by a binary origin for significant natal kicks \citep{Kuranov2009}, events captured by gravitational wave detectors require undisrupted binaries as progenitors.

In order for the natal kick not to disrupt the binary, the kick must be small enough and directed such that it opposes the orbit, resulting in a median velocity of 20~km/s provided to an undisrupted binary system \citep{Renzo2019}. 
This distribution was modelled by taking the velocity distribution provided by \cite{Renzo2019}'s fiducial simulation for NS+MS binaries and approximating it with a bimodal maxwellian with $w = 0.02, \sigma_1 = 1$~km/s and $\sigma_2 = 16$~km/s. While \cite{Renzo2019} focus only on the first supernova in a binary system, there is some evidence that second supernovae (assuming two massive components in a binary system) have smaller kicks \citep{Beniamini2016}. 
Proceeding here with the assumption that this will not significantly alter the velocity of the system, we ignore any second supernova kick in our modelling.
The distribution of undisrupted binaries shown in \autoref{fig:renzo} yields a spatial distribution much less puffed up than the isolated compact remnants. These binaries may be observed as X-ray binaries if they have sufficiently tight orbits with non-remnant stars \citep[e.g.][]{Hirai2021}, or by looking at orbital motions of stars with invisible companions \citep[]{Yamaguchi2018}.

\subsection{Can a nearby magnetar be the cause for rapid \texorpdfstring{$^{14}$}{14}C increases?}
The nearest magnetar to Earth is of great interest, for example because it could be the origin of rapid $^{14}$C increases which have been discovered in tree rings \citep{Wang2019}. Magnetars are thought to have spin-down timescales of a few thousand years and make up a large fraction of young pulsars \citep{Kaspi2017}. Assuming magnetars make up 50\% of the young pulsar population, our model finds that a magnetar is born every $\sim290$ years. This corresponds to around 34 magnetars formed across the Galaxy in the last $10\,000$ years. Assuming we observe up to 10~kpc (an approximation of our Galactic magnetar observation limits) we expect to detect half of these 34.
For comparison, the McGill Magnetar Catalog \citep{McGillCatalog} contains 22 confirmed magnetars in the Galaxy, with 6 further candidates, which --- given the uncertainties --- is in agreement with our model.
Using our model we estimate the local density of magnetars born in last 10,000 yrs to be $1.3\times 10^{-2}$ magnetars~kpc$^{-3}$ with an uncertainty of 6\% or $1.06\times 10^{-4}$ magnetars with heliocentric distance less than 200 pc. With this density we find the probably distance to the nearest magnetar to be 4.2~kpc. To improve the statistics we assume that the distribution of the age of magnetars is relatively constant for the last 1 Gyr. This shows that the estimated number of magnetars is too low, which makes magnetar as source for increase of $^{14}$C extraordinarily unlikely.

\begin{figure}
	\includegraphics[width=\columnwidth]{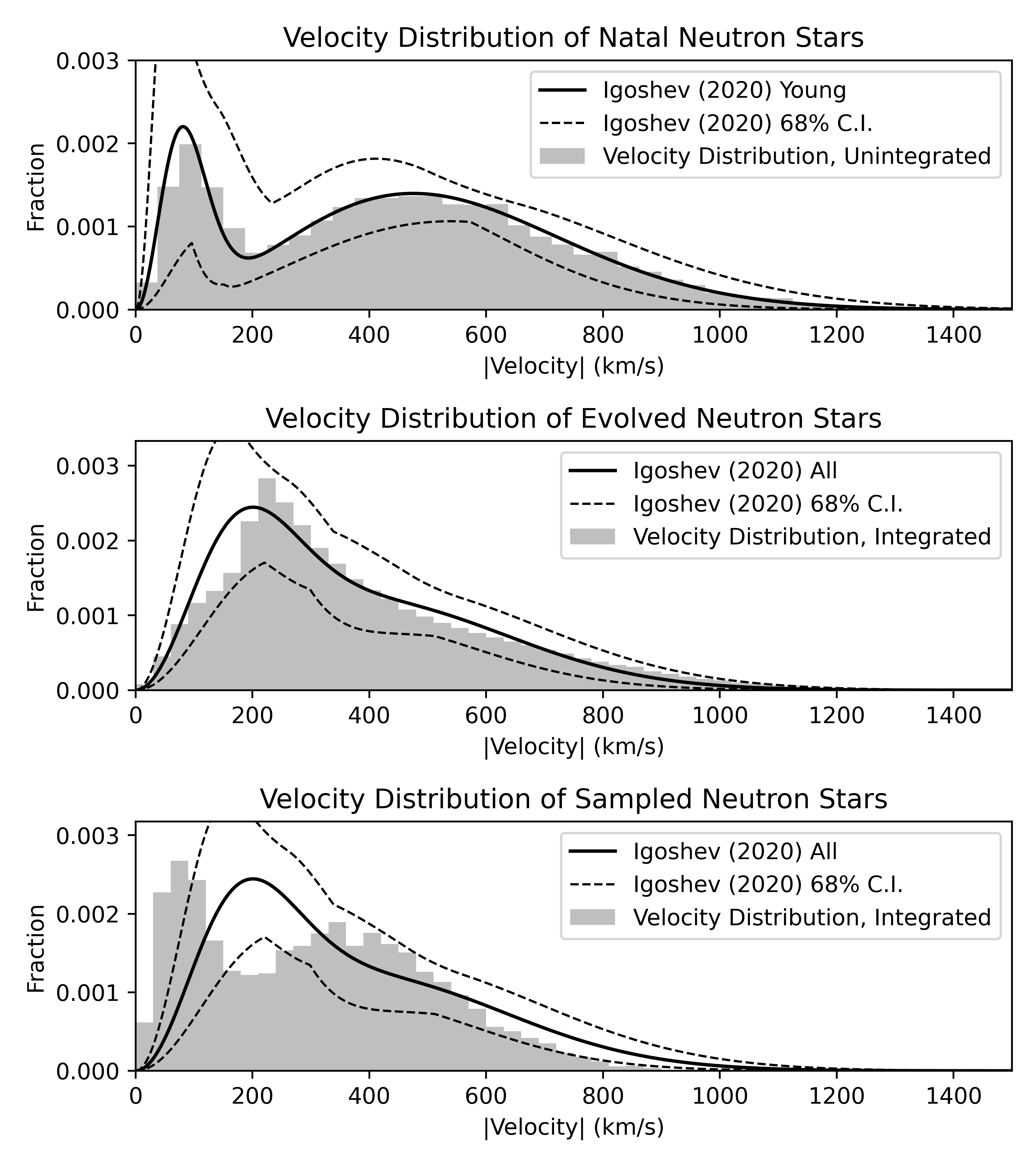}
    \caption{Peculiar velocity distribution of neutron stars. The empirical model of the observed velocity distribution (solid line), 
    based on isolated NSs with highly accurate astrometric measurements, 
    as presented in \citet{Igoshev2020} is compared to the velocity distributions found in this work (grey shaded region). The dashed lines are the 68\% confidence interval as stated by \citet{Igoshev2020}. The top panel is for young pulsars while the bottom two are for all observed pulsars.
    To emulate the selection function of observed pulsars, the remnants in our model were filtered to lie in $4<R/{\rm kpc}<16$ (where $R$ is the Galactic cylindrical radius) and galactic height $|z|< 8$~kpc, except for the middle panel which shows pulsars without such filtering.
    In the top panel the observed \citet{Igoshev2020} distribution is for young stars with spin down age less than 3 Myr. This is used to model the natal kicks in our model. The shaded histogram shows the distribution of peculiar velocities in our model after adding this kick to the intrinsic velocity of the stars which, as expected, matches the solid line. In our model, the distribution does not change much with the remnant age, hence to increase sample size we show remnants with age less than 2~Gyr. 
    The middle panel shows the velocity distribution of all pulsars in our dataset (without filtering) and it recovers the \citet{Igoshev2020} distribution for all measured pulsars, which is described by \autoref{eq:bimodal} with $w = 0.42, \sigma_1 = 128$~km/s and $\sigma_2 = 298$~km/s. The bottom panel also shows the velocity distribution of all pulsars, but with location filtering and remnant age less than 25 Myr (most pulsars from \citet{Igoshev2020} are younger than this). As before, to increase the sample size we consider objects with remnant age less than 2~Gyr and evolve them for 25~Myr.
    }
    \label{fig:hist}
\end{figure}

\subsection{Peculiar velocity distribution of neutron stars}

Confronting our Galactic underworld remnants distributions with direct observational constraint using contemporary data is not straightforward, particularly for the BH population. However, pulsars provide a way to probe the NS population.
The known pulsar NS population is of course heavily skewed toward young, radio-luminous pulsars which remain observable for only a moderately short time compared to the dynamical timescales of Galactic orbits. 
Observational pulsar data collated by \cite{Igoshev2020} demonstrated a noteworthy discrepancy between the peculiar velocity distribution of young (spin-down age <~3~Myr) and old NSs: young NSs exhibit two well-defined peaks in their distribution (top panel \autoref{fig:hist}) while a complete sample of NSs shows the peaks having merged (middle and bottom panels of \autoref{fig:hist}). 
Our {\tt GALAXIA} Galactic underworld model provides a straightforward dynamical explanation for the relaxation of the velocity distribution between young and old pulsars identified in the work by \citet{Igoshev2020}.
As dynamically perturbed remnant orbits are evolved in time through the Galactic potential the input natal distribution (\autoref{fig:hist} top panel) relaxes into a new distribution of NSs (\autoref{fig:hist} middle panel).
At first glance we might be encouraged: witnessing the total population of NSs relaxing into the velocity distribution of \citet{Igoshev2020}'s aged sample (\autoref{fig:hist} middle panel) for which it appears a good match. 
Unfortunately however, we do {\it not} recover the literature distribution if we further filter our population to emulate Igoshev's sampling: Galactocentric cylindrical radius $R$ between 4 and 16~kpc, Galactic height $z$ between -8 and 8~kpc and age <~25~Myr (\autoref{fig:hist} bottom panel). 
The filter has the effect of reducing the number of high velocity stars in the velocity distribution, caused by the Galactic $R$ and $|z|$ cuts, and reintroducing the electron-capture supernova peak. As can be seen in the bottom panel of \autoref{fig:hist}, the peculiar velocity distribution has undergone shape changes which make the fit worse than that of the complete population.
Therefore, the natal kick distributions from \citet{Igoshev2020} fails to evolve into the complete distribution when sampling effects are taken into account.
The prominence of the first peak could be reduced, and so a closer fit recovered, if NSs formed by electron-capture supernovae had shorter observable lifetimes. Equally, this could reveal that the sample of \citet{Igoshev2020} may have underlying biases, for example pulsars towards the Galactic centre may be under-represented.
It also should be noted that there is a known discrepancy between characteristic (spin-down) ages of pulsars and magnetars with their true ages by up to orders of magnitude due to effects such as magnetic field decay/growth \citep[e.g.][]{Kulkarni1986, Popov2012, Nakano2015}. Filtering out the NSs by their true age may not have the same effect as filtering by their characteristic ages.


\subsection{Escape of remnants}
Due to the natal kicks, many remnants exceed the Galactic escape velocity and therefore will eventually be ejected from the Galaxy. 
Escape velocity is location dependent so we compute it for the location of each remnant using {\tt galpy}.
We find that 30\% of remnants have escape velocity, or 40\% of NSs and 2\% of BHs. 
Integrating over Galactic history, we are therefore able to make a first estimate of the Galactic mass loss caused by the escape of compact remnants up to the present day, finding $2.1\times 10^8$~M$_{\sun}$, or $\sim 0.4\%$ of the present-day stellar mass of the Galaxy \citep[using a value of $5.04\times 10^{10}$~M$_{\sun}$;][]{cautun2020}. 
\cite{Olejak2020} have also 
explored the BHs using population synthesis modelling. Our predictions for total number of BHs as well as fraction of BHs escaping is broadly consistent
with theirs.  
We find around $8\times 10^7$ BHs in the Galaxy, while \cite{Olejak2020} found  $1.2\times 10^8$. We predict that  
2\% of BHs acquire escape velocity while \cite{Olejak2020} predict 5\%.

\subsection{Robustness of model assumptions}
Our present study ignores the effect of binary evolution. 
While beyond the scope of this manuscript, this could be incorporated into future modelling, building on example work such as that of \citet{Olejak2020}. However, we do not expect our results regarding the spatial distribution of remnants to be significantly affected by this. While NSs and BHs 
can grow in mass with time in some binary systems, they are all seeded by core collapse supernovae, which we account for. In addition to mass, the binary evolution affects the kick received by the remnants. 
A useful study on the topic by \cite{Renzo2019} found that $\sim$22\% of massive binaries merge prior to core collapse, becoming a single, massive star which evolves in isolation. The vast majority (77--97\%) of remaining binaries become unbound due to supernovae. For these systems the natal kick dominates the space velocity of pulsars, erasing effects of binary origin \citep{Kuranov2009}. Given that the kick distribution we use \citep{Igoshev2020} is sampled from a population of NSs that has no way to reject or distinguish disrupted binaries, any effects of binarity, however small, have been implicitly accounted for. 
About 10\% of systems remain bound after supernova and such systems can be probed by gravitational-wave interferometers.  
Such systems acquire a small kick from the supernovae, of order 20~km/s \citep{Renzo2019}. We model such systems separately and find that due to smaller natal kicks their spatial distribution is not as puffed up as isolated remnants, as shown in \autoref{fig:renzo}.

Remnant velocities will also be affected by remnant mass variances. As with binary evolution, the impact on the velocities of the remnants must be etched into the observed distribution of velocities. We have therefore accounted for these effects by using a distribution of velocities derived from observations of isolated NSs.

We tested the robustness of our resultant model spatial distribution by evolving the Galactic underworld with varying assumptions and initial conditions.
In particular, we trialled the alternate kick distribution provided by \citet{Hobbs2005}. 
The resulting distribution had the scale heights and lengths of each component (except for the scale length of BHs) with uncertainties overlapping those in \autoref{tab:scale_heights}, so our results appear robust even to significant changes in the underlying distributions.

\section{Conclusions}
In this paper, we have explored the spatial distribution and the kinematics of the Galactic underworld (compact remnants comprising NSs and BHs formed in supernovae terminating the lifecycles of massive stars) using a population synthesis model. The key advance over previous work is in the use of a population synthesis model that was designed to match the spatial distribution of the visible stars in the Galaxy and creates accurately tagged (in time, location and kinematics) stellar distributions that are correctly spawned based on galactic history. Additionally, we apply natal kicks to the remnants and evolve the perturbed population kinematically through the potential of the Galaxy. All of these factors play a crucial role in determining the final spatial distribution of the compact remnants. Our main findings are as follows.

\begin{itemize}
    \item The spatial distribution of compact remnants is different from that of visible stars. The remnants are more dispersed in the vertical direction with the scale height being about 3 times larger than that of the visible stars. This is mainly due to the significant velocity kicks received by the remnants at the time of their birth. 
    \item The spatial distribution of BHs is more centrally concentrated as compared to the NSs due to the smaller velocity kick they receive.
    \item  For some remnants the kick is so large that their total velocity becomes greater than their escape velocity (40\% of NS and 2\% of BHs). 
    We are able to estimate a Galactic mass loss in ejected compact remnants as $2.1 \times 10^8 {\rm M}_{\odot}$ or $\sim$0.4\% of the stellar mass of the Galaxy. 
    \item We explored the possibility of a magnetar being responsible for the rapid increase in $^{14}$C as discovered in the tree rings \citep{Wang2019}, but our population synthesis implies this is highly unlikely, requiring nearby objects not seen in our model.
    \item We find that our velocity distribution of pulsars evolves rapidly with time thereby providing a physical justification for different velocity distributions exhibited by old and young pulsars.
    While the general form for the relaxation in the distributions is in agreement, it was not possible to fully reproduce the observed distribution of the complete pulsar data in \citet{Igoshev2020}.
    We suggest that some unknown selection effects may be responsible for the residual misfit to the population.
\end{itemize}


Our results are built upon foundations laid down by a number of other researchers, notably the Galaxy population synthesis, natal kick distribution and the gravitational potential of the Milky Way. While we have here used the most up-to-date values available, inevitably new research will update these distributions and so alter the expected shape of the Galactic underworld. 
Foreshadowing this, the methods used in this paper have been kept deliberately general so that the public code made available with this work may be trivially updated to reflect new underlying assumptions or distributions.
A key outcome from this work is then the extension of {\tt GALAXIA} to now deliver accurate models of both the visible and invisible Galaxy.

Most BHs have so far been discovered in binary systems, which are strongly influenced by complex pathways of binary evolution. However, the recent detection of an isolated BH through microlensing \citep{Lam2022, Sahu2022} opens a new channel for probing the distribution of BHs in our Galaxy \citep{Wyrzykowski2022}. Mapping out the locations of isolated BHs with future observations will allow us to constrain the magnitude of BH kicks. This can in turn be used to progress our understanding of the evolution of binaries involving BHs, such as X-ray binaries and gravitational wave sources. 


\section*{Acknowledgements}
We would like to thank Richard Scalzo, Alberto Krone-Martins and Benjamin Pope for their helpful conversations.

We would also like to acknowledge the python packages used as part of our analysis: NumPy \citep{NumPy}, SciPy \citep{SciPy}, Matplotlib \citep{Matplotlib}, pandas \citep{pandas2010, pandas} and scikit-learn \citep{scikit-learn}.

The McGill Magnetar Catalog is maintained by the McGill Pulsar Group and is found at the following address: \url{http://www.physics.mcgill.ca/~pulsar/magnetar/main.html}. The Australian Telescope National Facility Pulsar Catalogue can be found at the following address: \url{http://www.atnf.csiro.au/research/pulsar/psrcat}.

\section*{Data Availability}

All data are available from the corresponding author upon reasonable request.

Our code is made available as part of the supplementary material.



\bibliographystyle{mnras}
\bibliography{example} 




\appendix

\section{Marginal distributions}
\label{sec:marginal_dist}

\begin{figure}
	\includegraphics[width=\columnwidth]{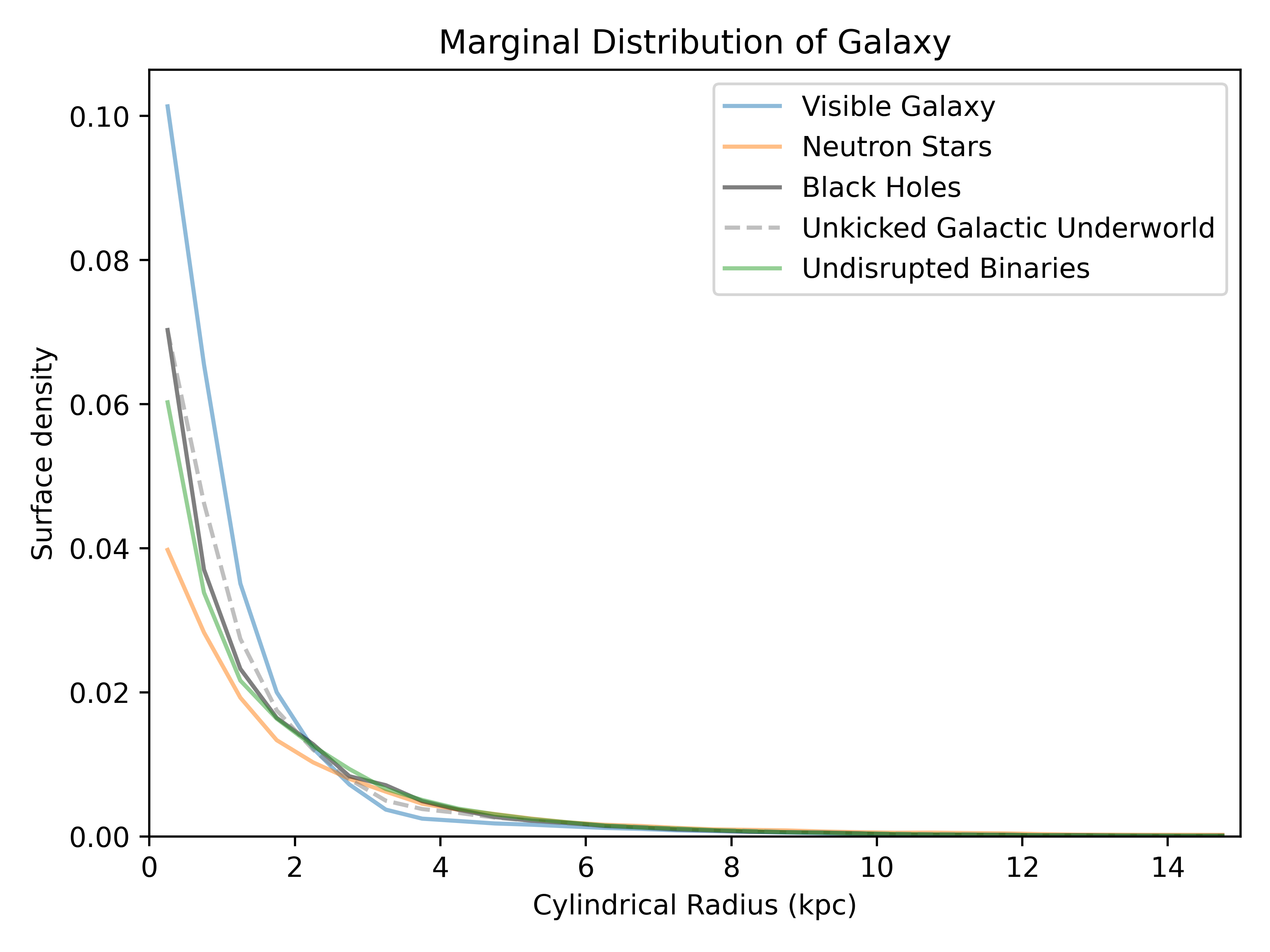}
	\includegraphics[width=\columnwidth]{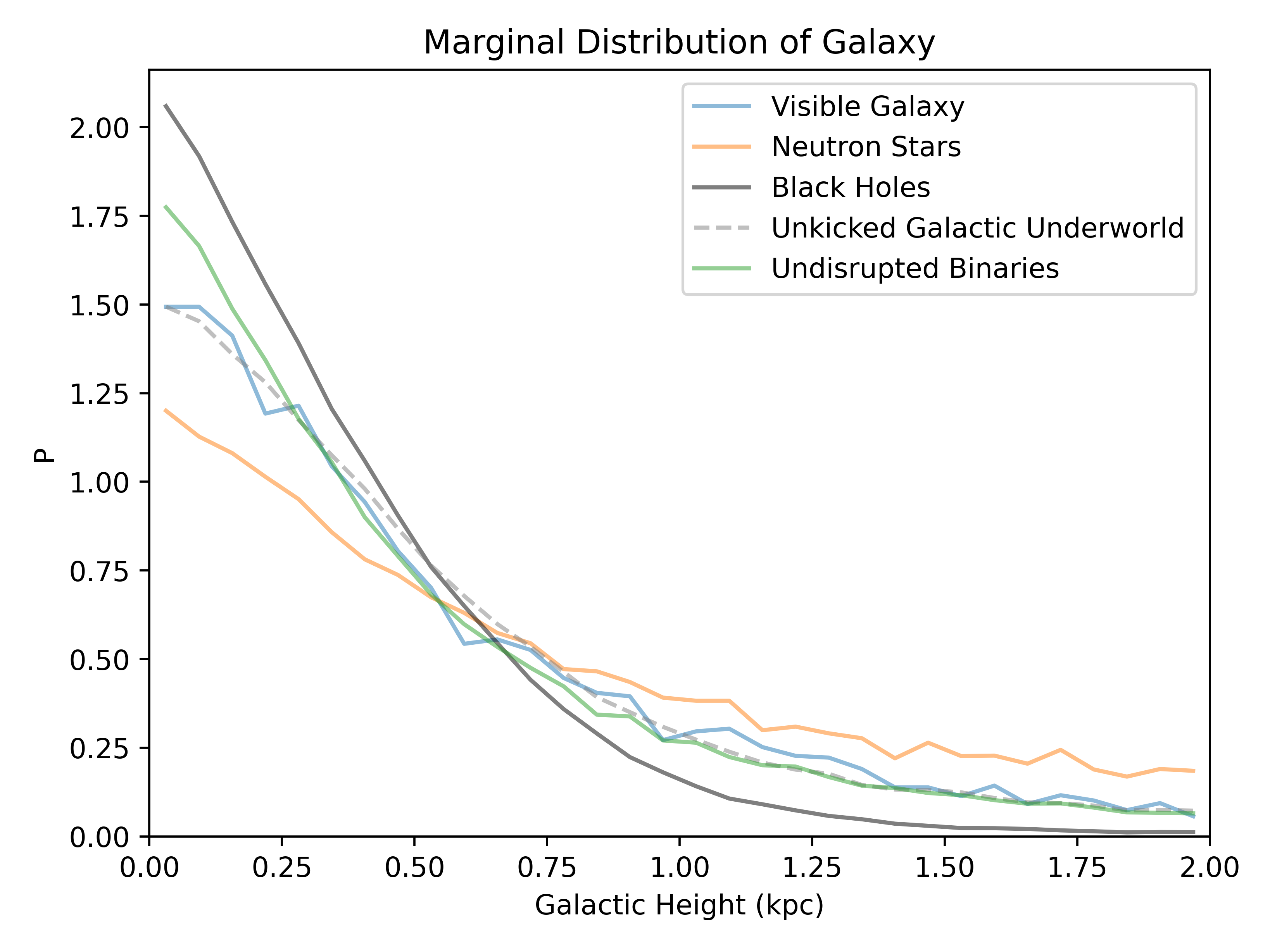}
    \caption{Marginal distributions of the visible Galaxy, NSs and BHs as a function of Galactocentric cylindrical radius and height above the Galactic plane. Both NSs and BHs are less centrally concentrated than the visible Galaxy, with BHs more centrally concentrated than NSs. These plots, when multiplied by their area, integrate to 1. The cylindrical radius plot appears not to however the area is increased due to the numbers stated being radii (and the relevant area being the inscribed circle's area).}
    \label{fig:marginal_dist}
\end{figure}

\begin{figure}
	\includegraphics[width=\columnwidth]{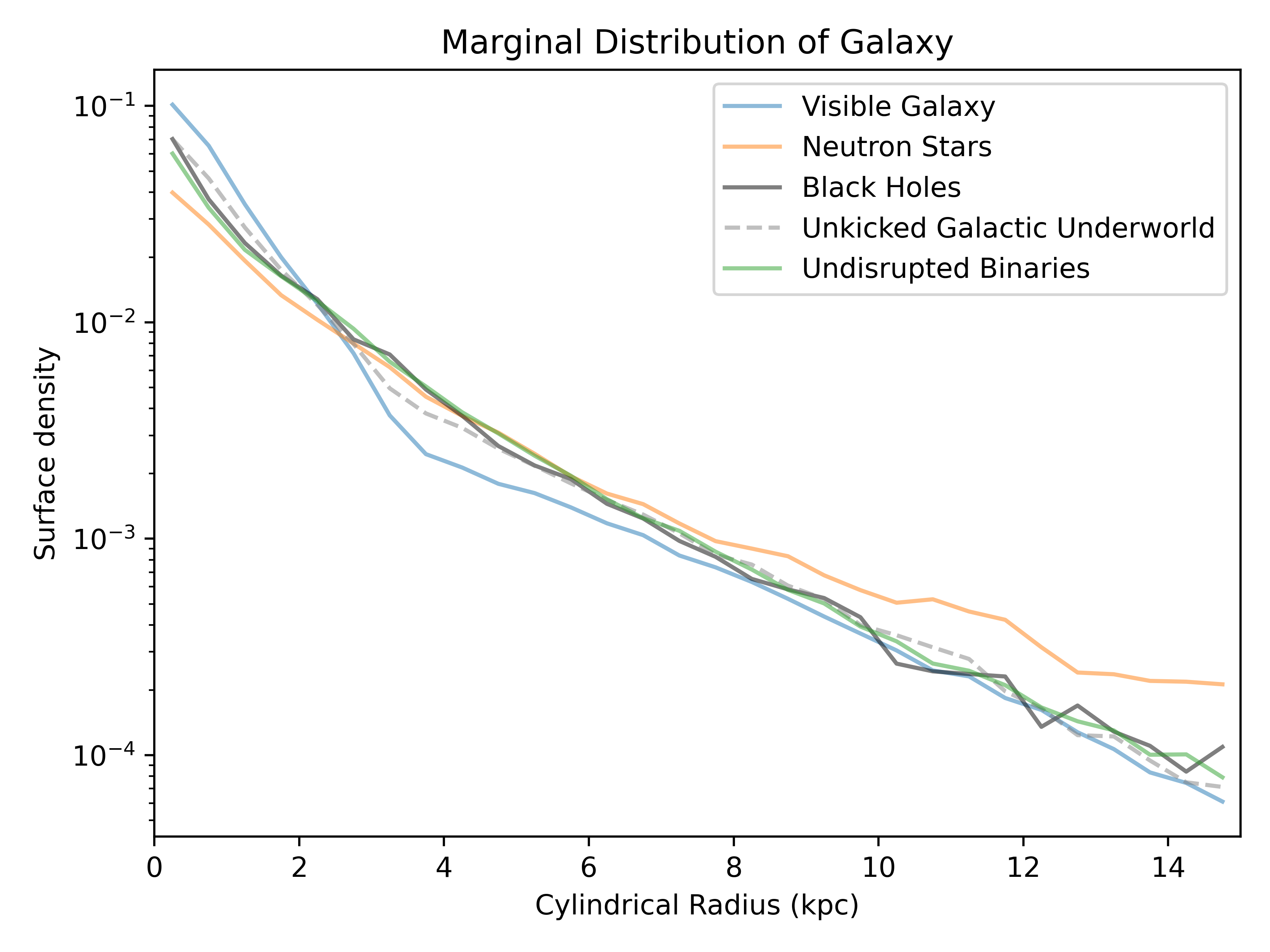}
	\includegraphics[width=\columnwidth]{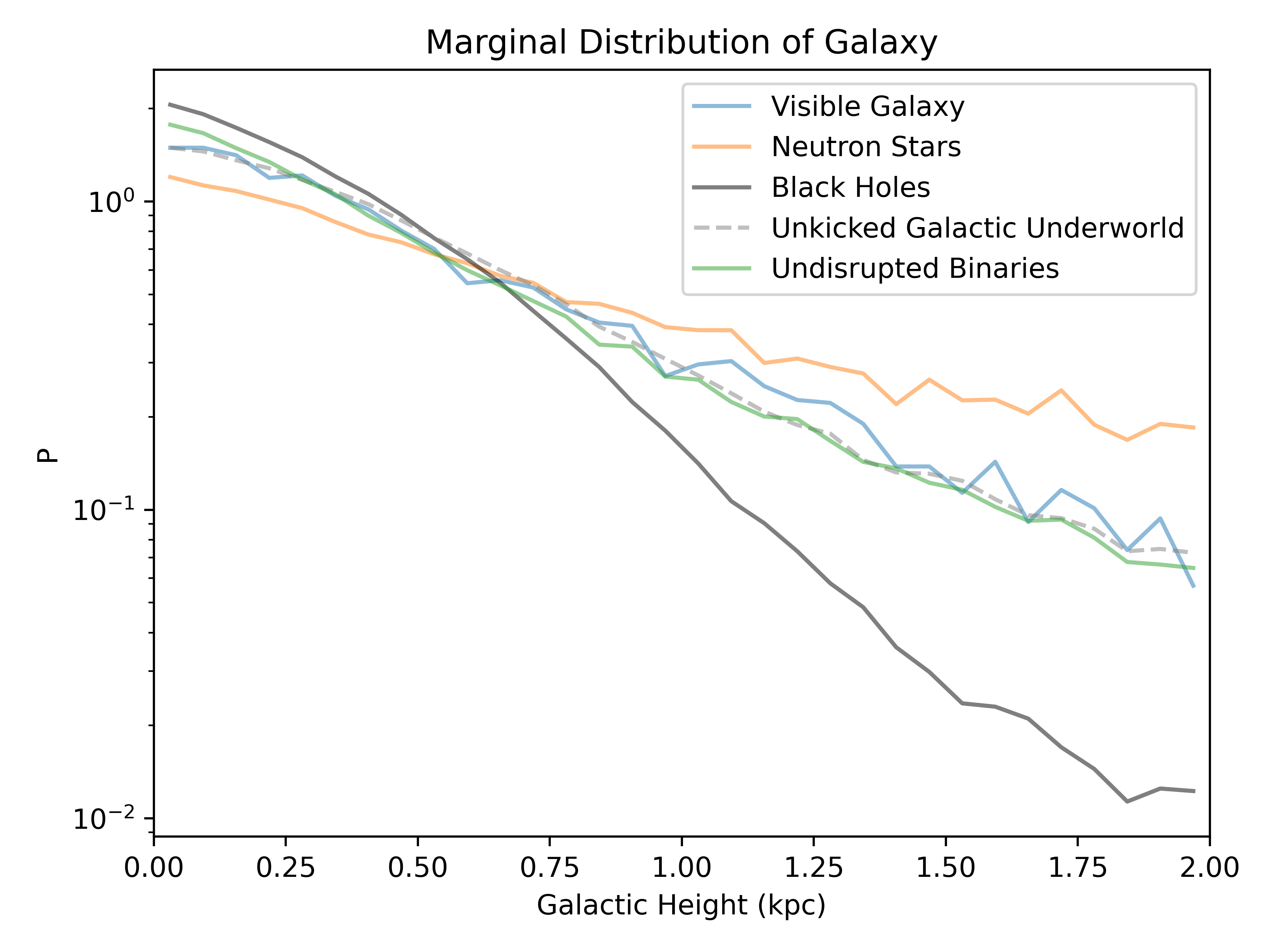}
    \caption{Marginal distributions of the visible Galaxy, NSs and BHs as a function of Galactocentric cylindrical radius and height above the Galactic plane with the y-axis on a log scale. Both NSs and BHs are less centrally concentrated than the visible Galaxy, with BHs more centrally concentrated than NSs.}
    \label{fig:log_marginal_dist}
\end{figure}

See \autoref{fig:marginal_dist} for the marginal distributions of the visible Galaxy, NSs and BHs as a function of Galactocentric cylindrical radius and height above the Galactic plane. The same plot is recreated in \autoref{fig:log_marginal_dist} but with the y-axis on a log scale.

\section{Unkicked Galactic Underworld}

\begin{figure}
	\includegraphics[width=\columnwidth]{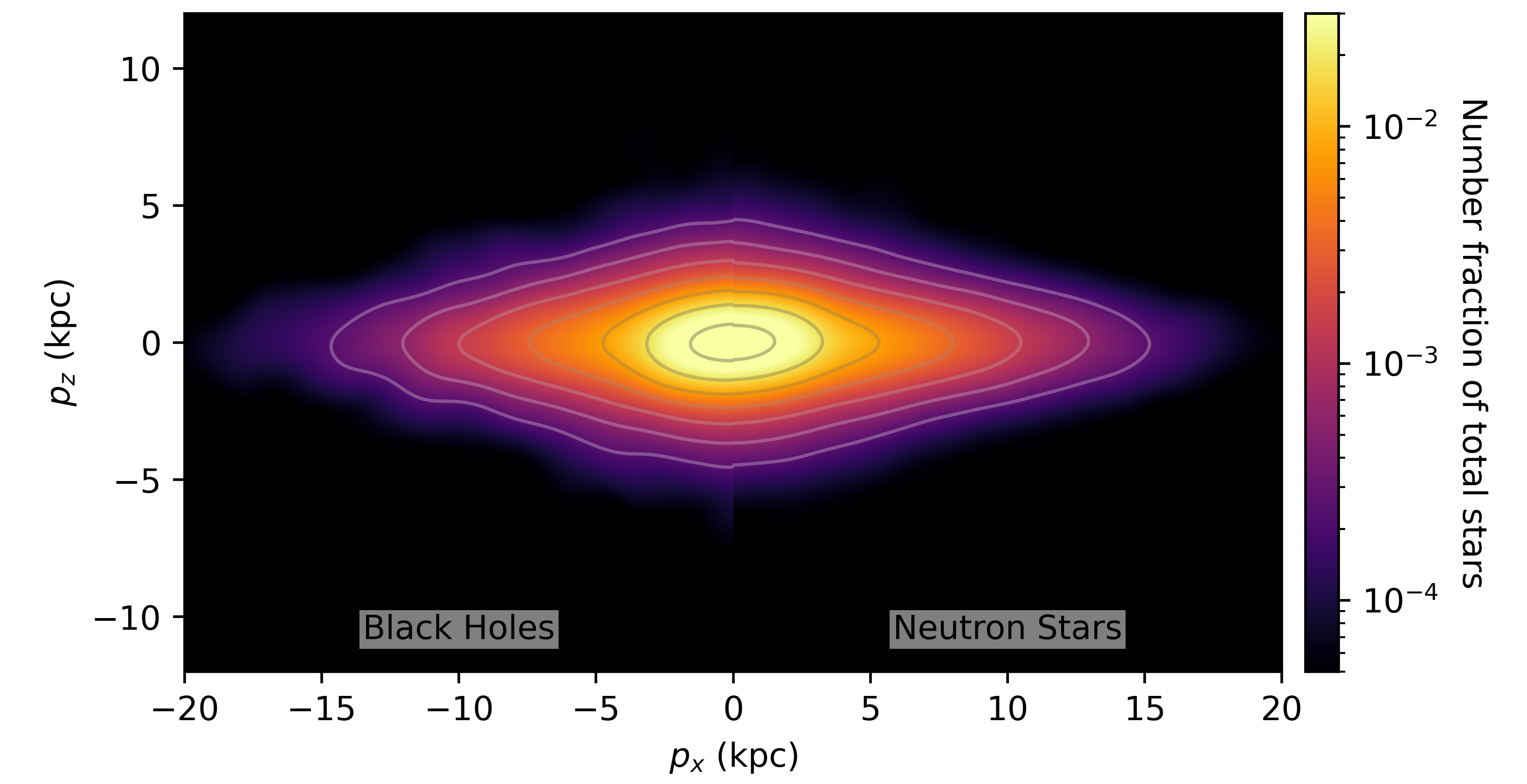}
    \caption{Side-by-side comparisons of unkicked BHs (left side) and NSs (right side). Logarithmically spaced contour lines are plotted on top. This plot was generated by passing the remnant locations through a gaussian kernel density estimator.}
    \label{fig:unkicked}
\end{figure}

See \autoref{fig:unkicked} for the distribution of the Galactic underworld if natal kicks are not applied.
\label{sec:unkicked}


\bsp	
\label{lastpage}
\end{document}